 \documentclass[final,3p,times,twocolumn,sort&compress]{elsarticle}

\usepackage{amsmath}
\usepackage{bm}        % for math]]
\usepackage{amsfonts}
\usepackage{flushend}

\usepackage{url}

\usepackage{color,xcolor}

\journal{Carbon}

\begin{document}

\begin{frontmatter}

\title{Emergence of Type-I and Type-II Dirac line nodes in penta-octa-graphene}

\author[mymainaddress,mysecondaryaddress,mythirdaddress]{Heng Gao}

\author[mymainaddress]{Wei Ren\corref{mycorrespondingauthor}}
\cortext[mycorrespondingauthor]{Corresponding author}
\ead{renwei@shu.edu.cn}

\address[mymainaddress]{International Centre for Quantum and Molecular Structures, Department of Physics, Shanghai University, Shanghai 200444, China}
\address[mysecondaryaddress]{Beijing National Laboratory for Condensed Matter Physics, Institute of Physics, Chinese Academy of Sciences, Beijing 100190, China}
\address[mythirdaddress]{Songshan Lake Materials Laboratory, Dongguan, Guangdong 523808, China}

\begin{abstract}
Carbon allotropes have a large family of materials with varieties of crystal structures and properties and can realize different topological phases. Using first principles calculations, we predict a new two-dimensional (2D) carbon allotrope, namely penta-octa-graphene, which consists of pentagonal and octagonal carbon rings. We find that penta-octa-graphene can host both type-I and type-II Dirac line nodes (DLNs). The band inversion between conduction and valence bands forms the type-I DLNs and the two highest valence bands form the type-II DLNs. We find that the type-I DLNs are robust to the biaxial strain and the type-II DLNs can be driven to type-I when applying over 3 $\%$ biaxial stretching strain. A lattice model based on the $\pi$ orbitals of carbons is derived to understand the coexistence mechanism of type-I and type-II DLNs in penta-octa-graphene. Possible realizations and characterizations of this penta-octa-graphene in the experiment are also discussed. Our findings shed new light on the study of the coexistence of multiple topological states in the 2D carbon allotropes.
\end{abstract}




\end{frontmatter}

%\linenumbers

\section{Introduction}
\label{}

In recent years, the interplay between topology and symmetry derives a rich variety of new topological quantum states in condensed matter physics. For instance, the time reversal symmetry protected topological insulators \cite{Hasan10p3045}, crystalline topological insulators \cite{Fu11p106802}, and crystal symmetry protected or enforced semimetals \cite{Gao19p153,Yong12p14045,Wang12p195320,Wang18p115164,Wieder16p186402,Weng15p011029} have been theoretically proposed and experimentally confirmed \cite{Xia09p398,Tanaka12p800,Liu14p864,Xu15p613}. These topological insulators and semimetals can host different kinds of novel topological quasi-particle fermions on their surface or in their bulk, like surface Dirac fermions \cite{zhang09p438} and hourglass fermions\cite{wang16p7598}, bulk Dirac and Weyl fermions \cite{Yong12p14045,Wan11p205101}, line nodes fermions \cite{Kim15p036806,Yu15p036807}, and three-component fermions \cite{Lv17p627}. The study of low energy topological quantum states in the condensed matter physics not only offers a platform to investigate the particles in high energy physics but also extends the new knowledge of quasi-particle fermions in solid \cite{Bradlyn16p5037}. In addition, the combination of crystal symmetries may allow the coexistence of multiple topological states in one solid state material. For instance, ZrTe can host both Weyl points and three-fold degenerate points \cite{Weng16p165201}, SrHgPb family materials possess both Dirac and Weyl points \cite{Gao18p106404}, and the coexistence is possible for Fermi arcs and Dirac cones on the surface in LaPtBi under in-plane compressive strain \cite{Lau17p076801}. These materials will offer opportunities to investigate the electronic and transport properties of topological multi-phases for the quantum spintronics.\par

Dirac line node (DLN) semimetals feature one-dimensional continuous Dirac points in the momentum space. It was proposed to realize the DLNs in the Cu$_3$PdN \cite{Kim15p036806}, CaAgX(X=P, As) \cite{Yamakage15p013708}, and ZrSiS family materials \cite{Schoop16p11696}. As analogue to the classification of type-I and type-II Dirac/Weyl cones according to their titled degree of cones, type-II DLNs semimetals have been theoretically proposed \cite{Li17p081106,Zhang17p4814}. Compared with the type-I nodal point, the cone of the type-II nodal point is completely tipped over so that there coexists electron and hole pockets at the certain energy level \cite{Soluyanov15p495}. Therefore the type-II DLNs can be considered as one-dimensional continuous type-II Dirac points. \par

Carbon allotropes consist of a large number of members and they vary quite different properties due to the intrinsic difference of crystal structures and chemical bonds. The first realized 2D material graphene \cite{Novoselov04p666} is the prototype of topological insulators \cite{Kane05p226801}, although the spin-orbital coupling (SOC) effect in graphene is too weak to observe the quantum spin Hall effect in experiment \cite{Yao07p041401}. Nonetheless, SOC-free of carbon element is an advantage to realize DLNs semimetals and Dirac semimetals in the carbon allotropes. It was proposed to realize DLNs in graphene network \cite{Weng15p045108}, 3D-C5 \cite{Zhong17p15641}, m-C8 \cite{Sung17pe361}, and bco-C16 \cite{Wang16p195501}, but so far all of them can only host one topological state of type-I DLNs.\par

In this work, by using first principles calculations and lattice model, we propose a new 2D carbon allotrope with pentagonal and octagonal carbon rings, named penta-octa-graphene, in which can coexist both type-I and type-II DLNs. The simulations of phonon dispersion and {\it ab initio} molecular dynamics reveal that penta-octa-graphene is dynamically and thermally stable. In this pent-octa-graphene, both type-I and type-II DLNs are symmetry-protected by in-plane glide mirror symmetry in the absence of SOC. The topological edge states from both type-I and type-II DLNs in the nanoribbons are investigated using the recursive Green’s function method. More interestingly, we find topological phase transition from the type-II to type-I of DLNs under the biaxial tensile strain. A tight binding model based on a square lattice with $\pi$ orbitals at octagonal carbon ring sites is proposed to understand the formation mechanism of type-II Dirac node line between two highest valence bands. Finally, we propose the periodical formation of 5-5-8 line defects in graphene to realize this penta-octa-graphene as well as characterizations of Raman spectra in experiment. \par

\section{Computational details}
We performed the structural optimization and electronic structure calculations within the framework of density functional theory (DFT) \cite{Tong66p1, Hohenberg64pB864} using Vienna {\it ab initio} simulation package (\textsc{VASP}) \cite{Kresse96p11169} based on the projector augmented wave (PAW) method \cite{Blochl94p17953}. The exchange-correlation interaction was treated within the generalized gradient approximation (GGA) \cite{Perdew96p3865} parametrized by Perdew, Burke, and Ernzerhof (PBE). The energy cutoff of 500\,eV was set in all the calculations and a Monkhorst-Pack grid with $11\times11\times1$ k-points was used for Brillouin zone integration. For the structural optimization, the lattice parameters and all the atoms are relaxed until the Hellmann-Feynman forces on all atoms are less than 0.005\,eV/\AA. The vacuum region of 10 $\rm{\AA}$ along z direction was included to simulate the monolayer penta-octa-graphene. The phonon dispersion was calculated using the finite displacement method \cite{Parlinski97p4063}, as implemented in the Phonopy code \cite{togo15p1}. The $\rm{p}_{z}$ orbitals of carbons at A and B sites were used to construct the maximum localization Wannier function (MLWF) using the \textsc{Wannier90} package\cite{mostofi08p685}. The edge states of penta-octa-graphene are calculated using the recursive Green{\textquoteright}s function method \cite{sancho85p851} from the obtained MLWF. \par

\section{Results and Discussions}
\begin{figure}[pt]
    \centering
    \includegraphics[width=8.6cm]{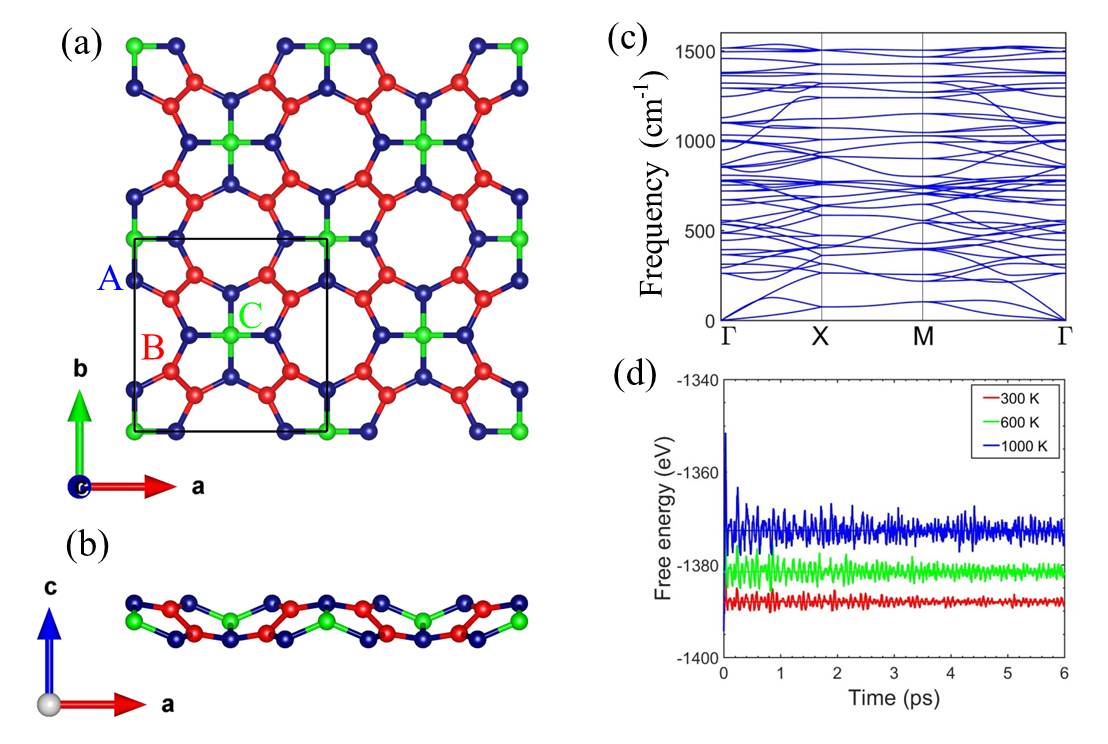}
    \caption{(a) Top view and (b) side view of the 2D layer structure of penta-octa-graphene. The blue, red, and yellow balls denote carbon atoms at A, B, and C sites. (c) Phonon dispersion of penta-octa-graphene. (d) Free energy evolution of $3\times3\times1$ penta-octa-graphene supercell at 300 K, 600 K, and 1000 K in 6 ps from the AIMD simulations. These horizontal lines indicate the average values of free energy at different temperatures.}
    \label{fig:fig_1}
\end{figure}

The structure of penta-octa-graphene is shown in Figure 1(a) and 1(b). In contrast to the hexagonal planar lattice of pristine graphene, the penta-octa-graphene features a square lattice with the lattice constant of 6.64 $\rm{\AA}$ and a buckling with the thickness of 1.24 $\rm{\AA}$. Other 2D carbon allotropes comprised of pentagonal and octagonal carbon rings have been theoretically proposed \cite{Su13p075453,Wang19p619,Wang18p6815}. However, the penta-octa-graphene has higher symmetry and can host DNLs due to the symmetry. It turns out that penta-octa-graphene has P$_4$/nmm (space group No. 129) symmetry which contains three symmetry generators that are $\{C_{4z}^{+}|\frac{1}{2}\frac{1}{2}0\}$, $\{C_{2x}|\frac{1}{2}\frac{1}{2}0\}$, and $\{I|\frac{1}{2}\frac{1}{2}0\}$. The nonsymmorphic operations of these symmetry operators play a crucial role to realize the type-II DLNs between the two highest valence bands. We will provide more information in the lattice model discussion later. Moreover, both type-I and type-II DNLs are protected by the in-plane glide plane symmetry operation $\{M_{z}|\frac{1}{2}\frac{1}{2}0\}$ which can be generated by the symmetry operations $\{C_{2x}|\frac{1}{2}\frac{1}{2}0\}$ and $\{I|\frac{1}{2}\frac{1}{2}0\}$. The unit cell of penta-octa-graphene is comprised of 8 pentagonal and 2 octagonal carbon rings and each octagonal carbon ring is adjacent to 8 pentagonal carbon rings. All carbon atoms in the unit cell can be categorized by three inequivalent carbon atoms at A, B, and C sites which are indicated by blue, red, and green colors in Figure 1(a). The bond distance between two B sites is $d_{BB}=1.44$ $\rm{\AA}$, the bonding distance of A and B sites is $d_{AB}=1.41$ $\rm{\AA}$, and the bonding distance of A and C sites is $d_{AC}=1.56$ $\rm{\AA}$. The geometry structure of penta-octa-graphene suggests that the A and B sites tend to form $sp^2$ bonding but C site carbons resemble $sp^3$ bonding. Compared with the full $sp^3$ in penta-graphene \cite{Zhang15p1372} and $sp^2$ in graphene, penta-octa-graphene is a 2D carbon allotrope with hybrid $sp^2$ and $sp^3$ states. \par

In order to prove that the penta-octa-graphene is energetic stable, we calculate the total energy and carbon density of the penta-octa-graphene as well as other 2D carbon allotropes (see Figure S1 in Supplementary Information). Although penta-octa-graphene is metastable compared with graphene, it is more stable than penta-graphene \cite{Zhang15p1372}, $\alpha$-graphyne, and $\beta$-graphyne \cite{Malko12p086804}. The carbon density of penta-octa-graphene is comparable to graphene and phagraphene \cite{Wang15p6182}. To further demonstrate the dynamical stability of penta-octa-graphene, the phonon dispersion of penta-octa-graphene is calculated using the finite displacement method \cite{Parlinski97p4063} and shown in Figure 1(c). One can see there is no imaginary frequency and it suggests that penta-octa-graphene is dynamically stable. To further verify the thermal stability of penta-octa-graphene, we preformed the {\it ab initio}  molecular dynamic simulations (AIMD) using NVT ensemble with the Nos\'e thermostat \cite{nose84p511}. The free energy of a $3 \times 3 \times 1$ penta-octa-graphene supercell with 162 carbon atoms in 6 ps at 300K, 600K, and 1000K was calculated and shown in Figure 1(d). We did not find disruption or structural reconstruction during the simulations, and the free energy fluctuations around certain values at different temperatures in Figure 1(d) also suggest penta-octa-graphene is thermally stable. To further verify the mechanical stability of penta-octa-graphene, the elastic constants are derived by fitting energy curves with respect to the uniaxial,biaxial, and shear strains (see Figure S2 in Supplementary Information). We have $C_{11}=C_{22}=141.7$ N/m, $C_{12}=62.4$ N/m, and $C_{66}=29.9$ N/m. The elastic constants are satisfied to the stability criteria $C_{11}C_{22}-C_{12}^{2}>0$ and $C_{66}>0$ \cite{Mouhat14p224104}. The Poisson's ratio of penta-octa-graphene $v_{12}=v_{21}=C_{12}/C_{11}=0.44$ is different from the negative Poisson's ratio in penta-graphene \cite{Zhang15p1372}.\par

\begin{figure}[pt]
    \centering
    \includegraphics[width=8cm]{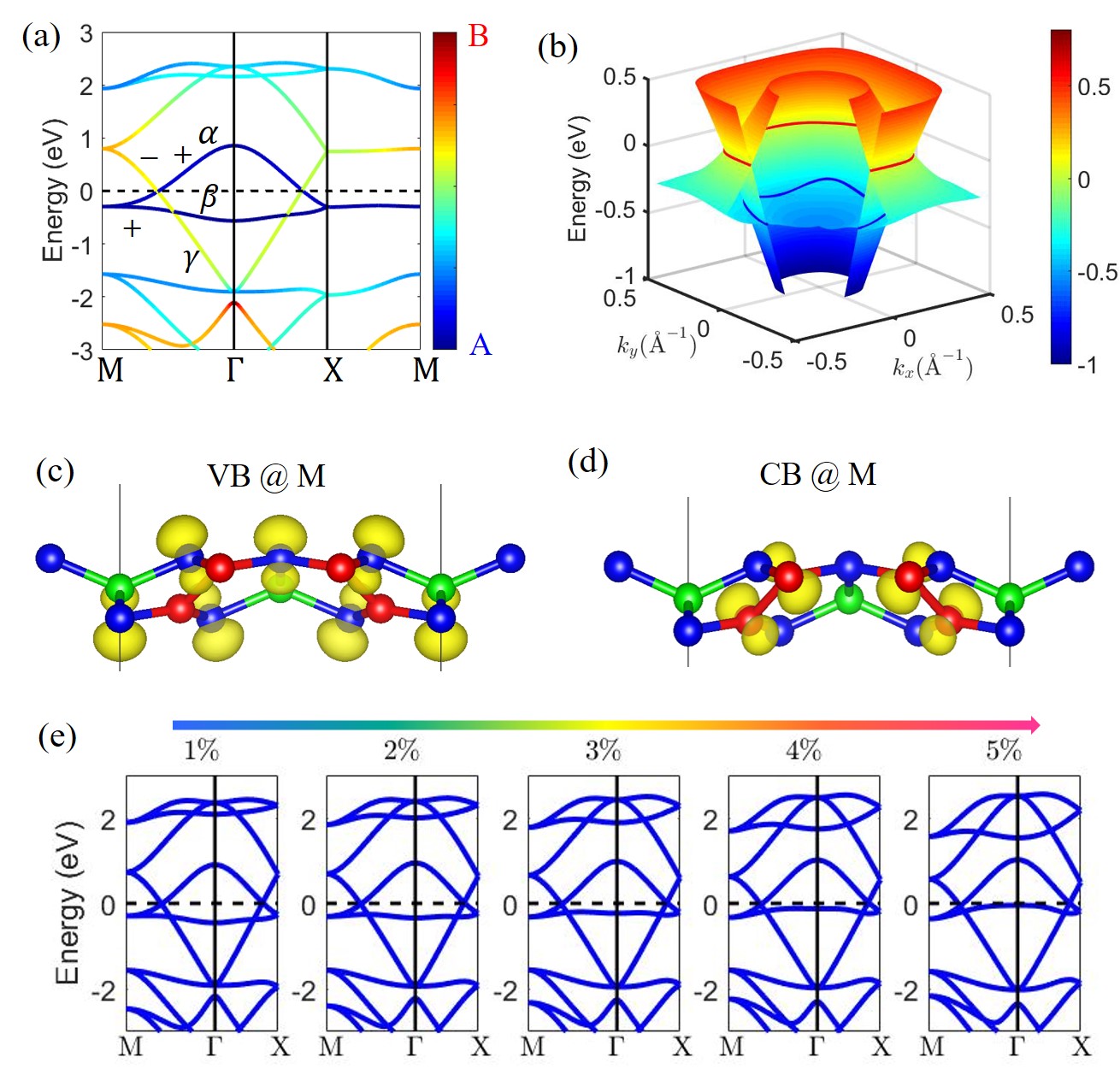}
    \caption{Electronic structures of penta-octa-graphene. (a)The blue and red colors denote the $p_{z}$ orbital contributions from carbon atoms at A and B sites, and the plus and minus signs indicate $M_{z}$ parity of each bands. (b) The red and blue lines in the 3D band structure denote the type-I and type-II DLNs, respectively. (c) and (d) show the real space wave functions of valence and conduction bands at M point. (e) provides the evolution of band structures of penta-octa-graphene under biaxial tensile strain from $1\%$ to $5\%$.}
    \label{fig:fig_2}
\end{figure} 

Figure 2(a) shows the band structure of penta-octa-graphene in the absence of SOC. There are two electron bands ($\beta$ and $\gamma$) and one hole band ($\alpha$) near the Fermi level. The band crossing points between the $\alpha$ and $\gamma$ bands along the $\rm{M}-\Gamma$ and $\Gamma-\rm{X}$ lines are type-I Dirac points. Besides, there exist type-II Dirac points near energy level -0.2 eV given by $\beta$ and $\gamma$ bands. To clearly show the coexistence of type-I and type-II DLNs, we plot the conduction band and two highest valence bands in three dimensions as shown in Figure 2(b). One can easily identity the type-I and type-II DLNs highlighted by red and blue lines, respectively. Both DLNs in penta-octa-graphene are symmetry-protected by the in-plane glide mirror symmetry $\{M_{z}|\frac{1}{2}\frac{1}{2}0\}$ which is generated by twice $\{C_{4z}^{+}|\frac{1}{2}\frac{1}{2}0\}$ and $\{I|\frac{1}{2}\frac{1}{2}0\}$ symmetry operations. $M_z$ parity of $\alpha$, $\beta$, and $\gamma$ bands without SOC was calculated and indicated by the plus and minus signs in Figure 2(a). The parity of $\gamma$ band is opposite to $\alpha$ and $\beta$ bands, such that the $\gamma$ band does not couple with $\alpha$ and $\beta$ bands and this is the reason why type-I and type-II DLNs in penta-octa-graphene remain gapless. In the presence of SOC, the mirror symmetry will not protect the DLNs any more since the double-degeneracy bands have opposite $M_{z}$ parity. To verify this point, we artificially increased the SOC strength to 100 times in the DFT simulations for the weak SOC in penta-octa-graphene. From the band structure (see Figure S3 in Supplementary Information), one can see both type-I and type-II DLNs are gapped by the SOC effect. Even we increase the SOC strength by 100 times, the gap along $\Gamma-M$ is still as small as 0.02 eV. The band gap opening mechanism is similar to the graphene and one should expect the Kane-Mele state in penta-octa-graphene. So the in-plane glide mirror symmetry can only protect the DLNs in the absence of SOC. \par

For a better understanding of DLNs in penta-octa-graphene, we calculate the orbital characteristics of band structure and real space wavefunctions of conduction and valence bands at M point. In Figure 2(a), (c), and (d), the results reveal that the $\alpha$, $\beta$, and $\gamma$ bands are dominated by the $p_{z}$ orbitals of carbon atoms at A and B sites. Similar to graphene, the band inversion between conduction and valence bands with $p_{z}$ orbitals forms the Dirac points. The Fermi velocities near Dirac points along $\Gamma-M$ and $\Gamma-X$ were calculated by the expression $v_{F}=E(k)/(\hbar k)$. It turns out that the Fermi velocities of $\gamma$ band near type-I Dirac points along the $\Gamma-M$ and $\Gamma-X$ are $6.7 \times 10^{5}$ m/s and $9.0 \times 10^{5}$ m/s which are comparable to graphene ($8.5 \times 10^{5}$ m/s) \cite{Wang12p590}. However, DLNs in penta-octa-graphene around the Fermi level should contribute more carriers than graphene. Figure 2(e) shows the evolution of band structures of penta-octa-graphene under biaxial tensile strain with $1\%-5\%$ range. The results show that type-I DLNs between $\alpha$ and $\gamma$ bands are robust to the biaxial strain which does not break any crystal symmetry. More interestingly, the type-II DLNs between $\beta$ and $\gamma$ bands are driven to type-I DLNs when the lattice of penta-octa-graphene is stretched over 3$\%$. We also calculated the dynamic stability of such stretched penta-octa-graphene. The phonon dispersion of penta-octa-graphene with biaxial $5\%$ tensile strain showed it was still dynamically stable (see Figure S4 in the Supplementary Information). The topological phase transition from type-II to type-I by biaxial tensile strain is thus feasible to be realized by lattice strain engineering in the experiment. \par

To investigate the edge states, we calculated band structures of semi-infinite penta-octa-graphene using the recursive Green{\textquoteright}s function method \cite{sancho85p851} based on the MLWF obtained from the $p_{z}$ orbitals of A and B sites. The edge states with zigzag and armchair edges are obtained and shown in Figure 3. It is known that the drumhead surface states are the hallmark of 3D DLNs semimetals. However, the 2D DLNs semimetals cannot host drumhead surface states, because the co-dimension of DLNs in 2D is 0 which is different from 1 for 3D case \cite{Zhao13p240404}. The DLNs in the 2D Brillouin zone has no projection along the direction perpendicular to it, so the 2D DLNs has no surface drumhead states like 3D case. In contrast to the surface drumhead states of 3D DLNs semimetal, the edge states of penta-octa-graphene are not topologically protected due to the 0 of co-dimension in 2D case. Comparing with the edge states in zigzag and armchair edges in Figure 3, one can find these edge states are edge dependent and induced by the trivial dangling bond. So the edge states can be tuned by the chemical functionalization on the edges by H or other atoms. \par

\begin{figure}[pt]
    \centering
    \includegraphics[width=8cm]{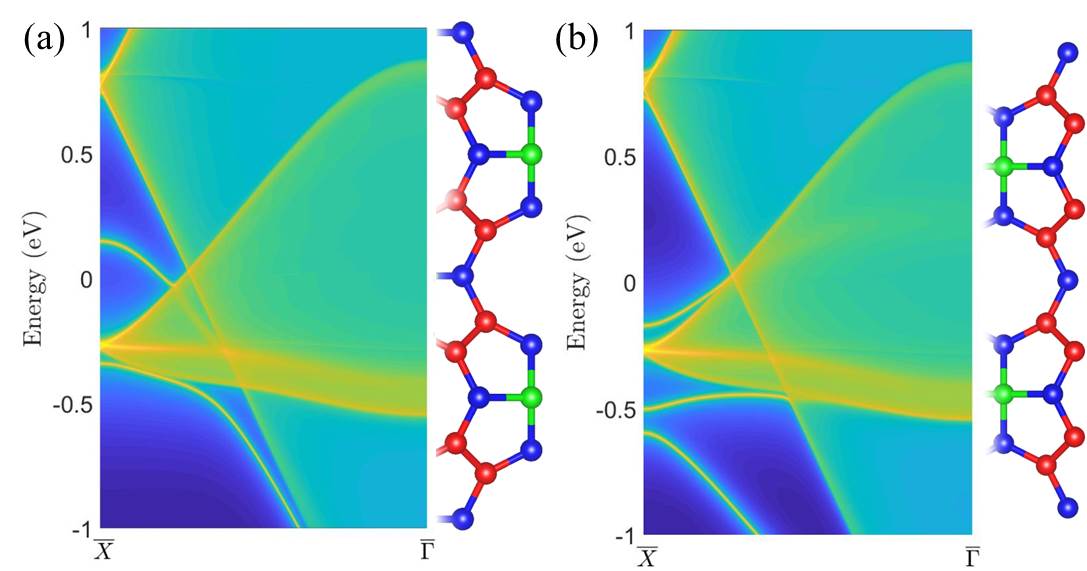}
    \caption{Edge states of penta-octa-graphene with (a) armchair edge (b)zigzag edge}
    \label{fig:fig_3}
\end{figure}

To understand the coexistence of Type-I and Type-II DLNs, we constructed a square lattice model based on the A and B sites as shown in Figure 4(a). As we pointed out, the conduction and valence bands are dominated by the $p_{z}$ orbitals of carbons at A and B sites. So we expect that the tight binding model for the lattice model with $p_{z}$ orbitals at A and B sites can reproduce the DLNs in DFT simulations. The tight binding Hamiltonian \cite{Porezag95p12947} with the nearest hopping reads as 
 \begin{equation}
\begin{split}
H=\epsilon_{1} \sum_{i}{ c_{A_{i}}^{\dagger}c_{A_{i}}}+\epsilon_{2}\sum_{j}{ c_{B_{j}}^{\dagger}c_{B_{j}}}+t_{1}\sum_{<ij>}{ c_{A_{i}}^{\dagger}c_{A_{j}}}\\
 +t_{2}\sum_{<ij>}{ c_{B_{i}}^{\dagger}c_{B_{j}}}+t_{3} \sum_{<ij>}{ c_{A_{i}}^{\dagger}c_{B_{j}}}
\end{split}
\end{equation}
where $c^{\dagger}$ and $c$ are the creation and annihilation operations, $\epsilon_{1}$ and $\epsilon_{2}$ are on-site energy for A and B sites, and $t_{1}$, $t_{2}$, and $t_3$ are the hopping parameters. The hopping parameters and on-site energies are obtained by fitting the DFT band structure. Then the eigenvalues along high-symmetry points are calculated by diagonalizing the $16 \times 16$ effective tight binding Hamiltonian matrix. As shown in Figure 4(c), the obtained band structure from the tight binding model can well reproduce both type-I and type-II DLNs. When we do not consider the nonsymmorphic symmetry operations in penta-octa-graphene and all the A and B sites are projected on the same plane, a smaller unit cell enables us to describe this symmorphic symmetry lattice model. The conventional cell with nonsymmorphic lattice can thus be considered as a $\sqrt{2}\times\sqrt{2}$ primitive cell with the symmorphic lattice. The corresponding Brillouin zones of nonsymmorphic and symmorphic lattices are denoted with black and purple lines in Figure 4(b). For the primitive cell with symmorphic lattice, we only need 4 A and 4 B sites and the band structure is calculated using the same hopping parameters and on-site energies in nonsymmorphic lattice as shown in Figure 4(d). We find that the Type-I DLNs between the conduction and valence bands still remain but the type-II DLNs disappear because of an unfolding process. \par

\begin{figure}[pt]
    \centering
    \includegraphics[width=8cm]{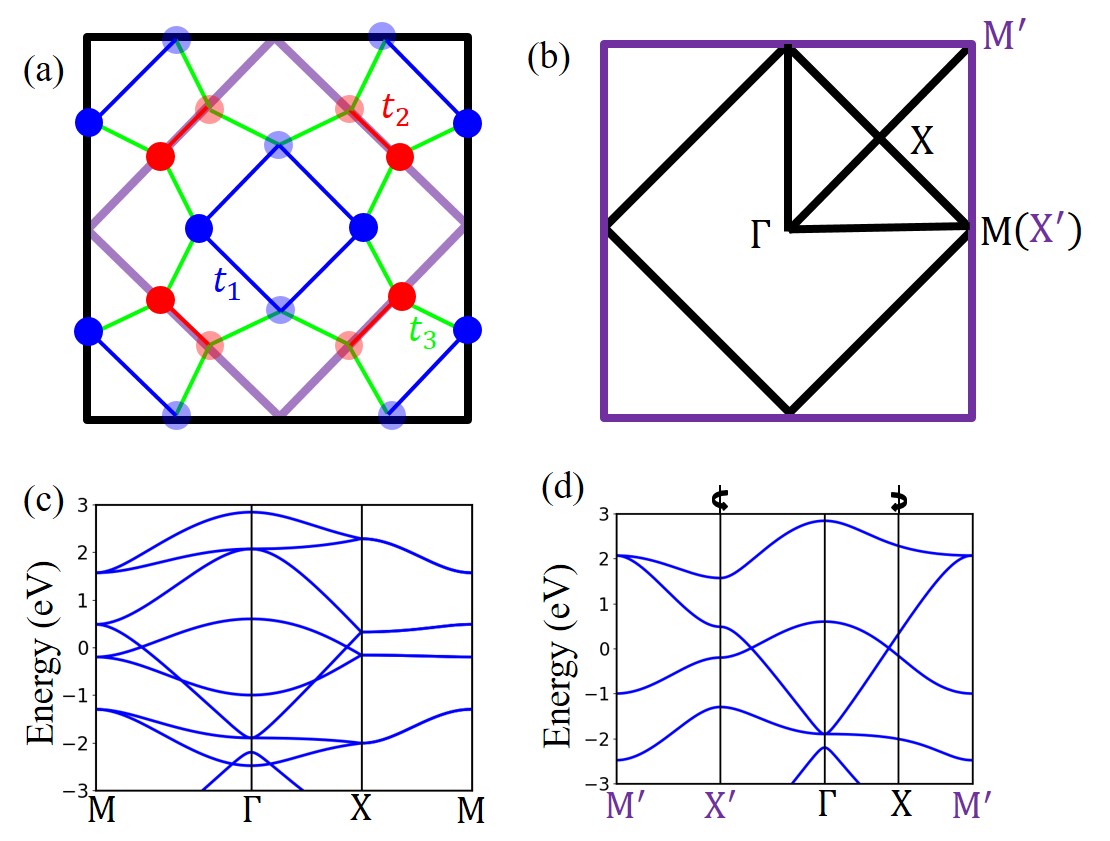}
    \caption{(a) Lattice model with A and B sites for penta-octa-graphene, the blue and red dots denote the A and B sites of carbons, respectively. The black and purple lines denote the unit cells of the nonsymmorphic and symmorphic lattices (b) Brillouin zones for the nonsymmorphic and symmorphic lattice with black and purple line, respectively. Band structures from lattice model for (c) nonsymmorphic and (d) symmorphic lattices. The hopping parameters are set to $t_{1}$=-0.4 eV, $t_{2}$=-3 eV, and $t_{3}$=-2.2 eV. The on-site energies of A and B sites are set to $\epsilon_{A}$= -0.2 eV and $\epsilon_{B}$=0.8 eV.}
    \label{fig:fig_4}
\end{figure}

Finally, we propose possible route for design and synthesis of penta-octa-graphene in the experiment. It was proposed that the approach of simultaneous electron irradiation and Joule heating by applied electric current could controllably produce 5-5-8 line defect in graphene \cite{Chen14p121407}. The 5-5-8 line defects observed in the experiment and the schematic drawing of the growth process of a 5-5-8 line defects in graphene are presented in Figure S5 in Supplementary Information. We noted that the 5-5-8 line defect is parallel to the direction of electric current. The atomic structure of penta-octa-graphene can be considered to be composed of the periodic 5-5-8 line defects. So the penta-octa-graphene would be produced when the generator of current is smoothly moving along the direction perpendicular to electric current. We expect that periodically fabricating 5-5-8 line defects in graphene using this technique will give rise to the penta-octa-graphene. Further, the Raman scattering spectrum is a powerful tool to characterize the carbon materials in the experiment \cite{Dresselhaus10p751}. In this work we have stimulated the Raman spectrum of penta-octa-graphene using density functional perturbation theory (DFPT) \cite{Refson06p155114}. It turns out that our penta-octa-graphene has 14 active Raman modes and the 3 strongest strength Raman modes are 301.5 $\rm{cm^{-1}}$, 655.8 $\rm{cm^{-1}}$, and 800.7 $\rm{cm^{-1}}$ (see Figure S6 in Supplementary Information). These distinct Raman signatures, like those of graphene and other carbon allotropes, can identity penta-octa-graphene in the experiment.\par

\section{Summary}
In summary, we proposed a new carbon allotrope to realize the coexistence of type-I and type-II DLNs. Compared with the features of Dirac points and $sp^{2}$ bonds in pristine graphene, our penta-octa-graphene has special DLNs and mixed $sp^{2}$ and $sp^{3}$ bonds. The dynamic and thermal stability of penta-octa-graphene has been verified and it is promising to be realized in the experiment. We also proposed the practical route for its synthesis, and characterization using Raman spectrum in the experiment. We have demonstrated that the biaxial tensile strain can drive the transition from type-II to type-I DLNs. The lattice model based on the $p_{z}$ orbitals at A and B sites can reproduce the band features and help understand the coexistence mechanism of type-I and type-II DLNs in penta-octa-graphene. The band inversion between conduction and valence bands forms the type-I DLNs, and the type-II DLNs come from the nonsymmorphic symmetry. Both type-I and type-II DLNs are protected by in-plane glide mirror symmetry in the absence of SOC. Although the artificially large SOC may break the DLNs, penta-octa-graphene is the ideal material for coexistence of the type-I and type-II DLNs. Our findings not only extend the family of carbon allotropes but also offer a new platform to realize multiple topological states in penta-octa-graphene. 

\section{Acknowledgments}
This work was supported by the National Natural Science Foundation of China (Grants No. 51672171, No. 51861145315, and No. 51911530124), the National Key Basic Research Program of China (Grant No. 2015CB921600), the fund of the State Key Laboratory of Solidification Processing in NWPU (SKLSP201703), the Austrian Research Promotion Agency (FFG, research Grant No. 870024, project acronym “MagnifiSens”). The Department for Integrated Sensor Systems also gratefully acknowledges partial financial support by the European Regional Development Fund (EFRE) and the province of Lower Austria. The supercomputing services from AM-HPC and Shanghai Supercomputer Center are also acknowledged. H.G. acknowledges support from National Postdoctoral Program for Innovative Talents (Grant No. BX20190361).

\bibliographystyle{elsarticle-num}
\bibliography{ref}

\end{document}